\documentclass[oneside,a4paper,11pt,reqno]{amsart}
\usepackage{amsmath, amstext,amsbsy,amsopn,amsthm, amscd,amsfonts,amssymb}
\usepackage{graphics,epsfig}
\usepackage[usenames]{color}
\usepackage[round]{natbib}
\usepackage[colorlinks=true,citecolor= blue,linkcolor=blue]{hyperref}
\newcommand{\begineq}[1]{\begin{equation}\label{#1}}

\newcommand{\eqend}{\end{equation}}

\newcommand{\eg}{\textit{e.g.}}

\newcommand{\refeq}[1]{Eq.\ (\ref{#1})}

\newcommand{\dif}{\text{d}}

\newcommand{\stressN}{\mathbf{S}}

\newcommand{\stress}{T}
\newcommand{\defgT}{\mathbf{A}}

\newcommand{\IIT}{\mathbf{I}}

\newcommand{\rank}[1]{({#1})}

\newcommand{\volfrac}{c}

\newcommand{\Trace}{\text{Tr}}

\begin{document} 
\date{\today}
\title[Neo-Hookean fiber composites undergoing finite out-of-plane shear deformations]
{Neo-Hookean fiber composites undergoing finite out-of-plane shear deformations}
\author{G.\ de{B}otton$^*$ and I.\ Hariton}
\maketitle
\centerline{The Pearlstone Center for Aeronautical Studies}
\centerline{Department of Mechanical Engineering}
\centerline{Ben-Gurion University}
\centerline{Beer-Sheva 84105}
\centerline{P.O.Box 653}
\centerline{Israel}
%
%
\begin{abstract}
The response of a neo-Hookean fiber composite undergoing finite out-of-plane shear deformation is examined. 
To this end an explicit close form solution for the out-of-plane shear response of a cylindrical composite element is introduced.
We find that the overall response of the cylindrical composite element can be characterized by a fictitious homogeneous neo-Hookean material.
Accordingly, this macroscopic response is identical to the response of a composite cylinder assemblage.
The expression for the effective shear modulus of the composite cylinder assemblage is identical to the corresponding expression in the limit of small deformation elasticity, and hence also to the expression for the Hashin-Shtrikman bounds on the out-of-plane shear modulus.
\newline
\newline \textbf{keywords: fiber composites, hyperelastic composites, nonlinear composites, finite elasticity, effective properties, micromechanics}  
\end{abstract}
%
\section{introduction}\label{section1}
\setcounter{footnote}{1}
\footnotetext{{
E-mail: debotton@bgumail.bgu.ac.il, Tel: int. (972) 8 - 647 7105}}
The problem of characterizing the behavior of fiber reinforced solids undergoing large deformations arises in many modern applications such as biomechanics and active materials.
A desirable goal is to express the behavior of these materials in terms of the properties and the volume fractions of the constituents composing the composite.
This will allow, for example, to account for variations in the volume fractions of the phases and to obtain estimates for the stresses developing in the constituents.
Motivated by this goal, in this work we study the problem of characterizing the behavior of fiber reinforced composites under out-of-plane shear loading.
This type of loading may arise, for instance, during a torsion of a cylinder made out of a composite reinforced with circumferential fibers. 

Following the work of \citet{hash&rose64jamt}, who considered the analogous problem in the limit of small deformation elasticity, we examine the problem of a single \emph{cylindrical composite element} \citep[\eg,][]{hill64jmps} subjected to shear loading in the fiber direction.
For the case of a composite element made out of two incompressible neo-Hookean phases we provide a simple close form solution describing the strains and the stresses in the two phases.
These expressions correctly reduce to the small deformation limit and are in agreement with the results of \citet{hash&rose64jamt}.
The solution can be easily extended to coated composite elements, that is the case where the fibers are coated with a few concentric shells made out of different materials.

We further demonstrate that, under this type of loading, the behavior of the composite element is identical to that of a fictitious homogeneous neo-Hookean material.
An expression for the \emph{effective} shear modulus of this fictitious material, in terms of the properties and the volume fractions of the phases composing the composite element, is determined.
This expression is identical to the corresponding expression of \citet{hash&rose64jamt} for the out-of-plane shear modulus of a composite cylinder assemblage in the small deformation limit.
By following arguments similar to those of \citet{hash&rose64jamt}, we also conclude that this shear modulus describes the macroscopic behavior of a {composite cylinder assemblage} undergoing finite out-of-plane shear deformations.
Finally, we recall that in the limit of small deformation elasticity the expression for the effective out-of-plane shear modulus is identical to the expression for the Hashin-Shtrikman bounds of \citet{hash65jmps} on the out-of-plane shear modulus of two phase transversely isotropic composites.

%
\section{Out-of-plane shear deformation of a cylindrical composite element}\label{ASLC}
We consider a {cylindrical composite element} made out of a fiber with a circular cross section surrounded by a concentric shell of a different material. 
The elastic behaviors of both phases are characterized by neo-Hookean strain energy-density functions
\begineq{W neo hookean}
W^{\rank{s}}(\defgT)=\textstyle{\frac{1}{2}}\mu^{\rank{s}}\Trace{\left(\defgT^{T}\defgT-\IIT\right)},
\eqend
where $s=1$ for the fiber phase and $s=2$ for the shell.
The radius of the fiber-shell interface is $a$, the outer radius of the shell is $b$, and we assume a perfect bonding at the interface between the two phases.
It is assumed that the diameter of the cylindrical element is at least an order of magnitude smaller than its length.
We choose a coordinate system such that in the reference configuration the $X_{3}$ axis is aligned with the direction of the fiber. 
The composite element is subjected to a simple out-of-plane shear deformation such that in the deformed configuration the location of the points on the outer boundary of the shell is given by the mapping 
\begineq{boundary cartesian mapping}
x_{1}=X_{1},\quad
x_{2}=X_{2},\quad
x_{3}=X_{3}+\gamma X_{1},
\eqend
where $\gamma$ is the amount of shear \citep{ogden97book}.
In a cylindrical coordinate system, where the $Z$ axis is aligned with the fiber direction and the ray $\Theta=0$ with the positive direction of the $X_{1}$ axis, the above mapping is expressed in the form
\begineq{boundary cylindrical mapping}
r=R,\quad
\theta =\Theta,\quad
z=Z+\gamma R\cos\Theta,
\eqend
where $(R, \Theta,Z)$ and $(r,\theta,z)$ are the coordinates of the boundary points in the reference and the deformed configurations, respectively.

For the material points in the two phases we consider the following mappings  
\begineq{cylindrical mapping}
r^{\rank{s}}=R^{\rank{s}},\quad
\theta^{\rank{s}} =\Theta^{\rank{s}},\quad
z^{\rank{s}}=Z^{\rank{s}}+\gamma f^{\rank{s}}(R^{\rank{s}})\cos\Theta^{\rank{s}},
\eqend
where the functions $f^{\rank{s}}$ are functions of $R^{\rank{s}}$ only.
The continuity condition across the interface resulting from the third of \refeq{cylindrical mapping} is
\begineq{disp interface cond}
f^{\rank{1}}(a)=f^{\rank{2}}(a),
\eqend
and at the outer boundary of the shell the boundary condition results in the requirement
\begineq{disp boundary}
f^{\rank{2}}(b)=b.
\eqend

In cylindrical coordinates the expressions for the deformation gradients in the two phases are \citep[\eg,][]{hump02book},
\begineq{polar def grads}
\defgT^{\rank{s}} = 
\begin{pmatrix}
     1 & 0 & 0   \\
      0 & 1 & 0 \\
     \gamma f_{R}^{\rank{s}}\cos\Theta^{\rank{s}} & 
     -\frac{1}{R}\gamma f^{\rank{s}}\sin\Theta^{\rank{s}} & 1
\end{pmatrix}
\eqend
where $f_{R}$ is the derivative of $f$ with respect to its argument. We note that $\det{\defgT^{\rank{s}}}=1$.

In terms of $\defgT^{\rank{s}}$, the nominal (or Piola-Kirchhoff) stresses in the two neo-Hookean phases are
\begineq{polar nominal stress}
\stressN^{\rank{s}}=
\begin{pmatrix}
   \mu^{\rank{s}}-p^{\rank{s}} & 0 & \gamma p^{\rank{s}} f_{R}^{\rank{s}}\cos\Theta^{\rank{s}}   \\
  0 & \mu^{\rank{s}}-p^{\rank{s}} & -\frac{1}{R}\gamma p^{\rank{s}} f^{\rank{s}}\sin\Theta^{\rank{s}} \\
     \gamma  \mu^{\rank{s}} f_{R}^{\rank{s}}\cos\Theta^{\rank{s}} & 
     -\frac{1}{R}\gamma  \mu^{\rank{s}} f^{\rank{s}}\sin\Theta^{\rank{s}} &  \mu^{\rank{s}}-p^{\rank{s}}
\end{pmatrix}
\eqend
where $\mu^{\rank{s}}$ is the shear modulus of the $s$-phase and $p^{\rank{s}}$ is the arbitrary pressure in the phase.
In accordance with the assumption that the length of the cylindrical composite element is substantially longer than its diameter, far from the bases the pressure is independent of the $Z$ coordinate, that is $p^{\rank{s}}=p^{\rank{s}}(R,\Theta)$. 
The traction continuity conditions across the interface between the two phases are
\begineq{traction at the interface}
\begin{split}
\mu^{\rank{1}}-p^{\rank{1}}(a,\Theta)&=\mu^{\rank{2}}-p^{\rank{2}}(a,\Theta),\\
\mu^{\rank{1}} f_{R}^{\rank{1}}(a) &=
\mu^{\rank{2}} f_{R}^{\rank{2}}(a),
\end{split}
\eqend
and we note that the traction continuity condition for $S_{R \Theta}$ is satisfied.

For convenience, we follow \citet{ogden97book} and express the equilibrium equations in terms of the nominal stress. 
Assuming no body forces, in cylindrical coordinates \citep[\eg,][]{hump02book} these equations lead to the following three equations, namely,
\begineq{equilibrium equations}
\frac{\partial p^{\rank{s}}}{\partial R}=0,\quad
\frac{1}{R}\frac{\partial p^{\rank{s}}}{\partial \Theta}=0,\quad
{-\frac{1}{R^{2}}f^{\rank{s}}+{\frac{1}{R}\frac{\dif f^{\rank{s}}}{\dif R}+\frac{\dif^{2} f^{\rank{s}}}{\dif R^{2}}}}
=0.
\eqend
From the equilibrium along the $R$ and $\Theta$ directions it follows that the pressure in each phase is constant,
and from the equilibrium along the $Z$ direction
\begineq{form of f}
f^{\rank{s}}(R)=C^{\rank{s}}_{1}R+\frac{C^{\rank{s}}_{2}}{R}.
\eqend

We complete the solution by noting that in the fiber phase we must choose $C^{\rank{1}}_{2}=0$.
The remaining three integration constants in \refeq{form of f} are determined from Eqs.~(\ref{disp interface cond}), (\ref{disp boundary}) and the second of (\ref{traction at the interface}).
Accordingly, in the fiber phase
\begineq{z mapping fiber}
z=Z+\gamma \frac{2 \mu^{\rank{2}}}
{(1-\volfrac^{\rank{1}})\mu^{\rank{1}}+(1+\volfrac^{\rank{1}})\mu^{\rank{2}} }
R\cos\Theta,
\eqend
and in the shell
\begineq{z mapping shell}
z=Z+\gamma
 \frac{(\mu^{\rank{1}}+\mu^{\rank{2}}) - \volfrac^{\rank{1}}({\frac{b}{R}})^{2}(\mu^{\rank{1}}-\mu^{\rank{2}})}
{(1-\volfrac^{\rank{1}})\mu^{\rank{1}}+(1+\volfrac^{\rank{1}})\mu^{\rank{2}} }
R\cos\Theta,
\eqend
where $\volfrac^{\rank{1}}=(a/b)^{2}$ is the volume fraction of the fiber phase in the composite element.
The jump in the pressure is determined from the first of \refeq{traction at the interface}.
To determine the absolute values of the pressure additional information concerning the stresses must be specified.
Finally, the explicit expressions for the stresses in the two phases can be readily determined.

\section{Comparison with a finite element simulation}

We compare the analytic solution developed in the previous section with corresponding numerical finite element simulation.
Taking into account symmetry conditions, a quarter of the cylindrical composite element $(0\le\Theta\le\pi/2)$ was constructed and meshed by application of the commercial finite element code ABAQUS.
The ratio of the length of the cylindrical composite element to its diameter was 5.
The volume fraction of the fiber was $\volfrac^{\rank{1}}=0.25$ corresponding to $a/b=2$.
The nodes in the $(R,Z)$-plane containing the ray $\Theta=0$ were constrained from moving in the $\Theta$ direction.
The nodes in the $(R,Z)$-plane containing the ray $\Theta=\pi/2$ were constrained from moving in the $\Theta$ and the $Z$ directions.
The boundary condition according to \refeq{boundary cartesian mapping} for the circumferential nodes were setup manually for $\gamma=0.4$.
This value of $\gamma$ corresponds to a principal stretch ratio $\lambda=1.22$ \citep[\eg,][]{ogden97book}.
ABAQUS built in neo-Hookean constitutive model was used and the ratio between the shear moduli of the fiber phase and the shell was $\mu^{\rank{1}}/\mu^{\rank{2}}=10$.

\begin{figure}[t]
\begin{center}
\includegraphics[width=1.0\textwidth]{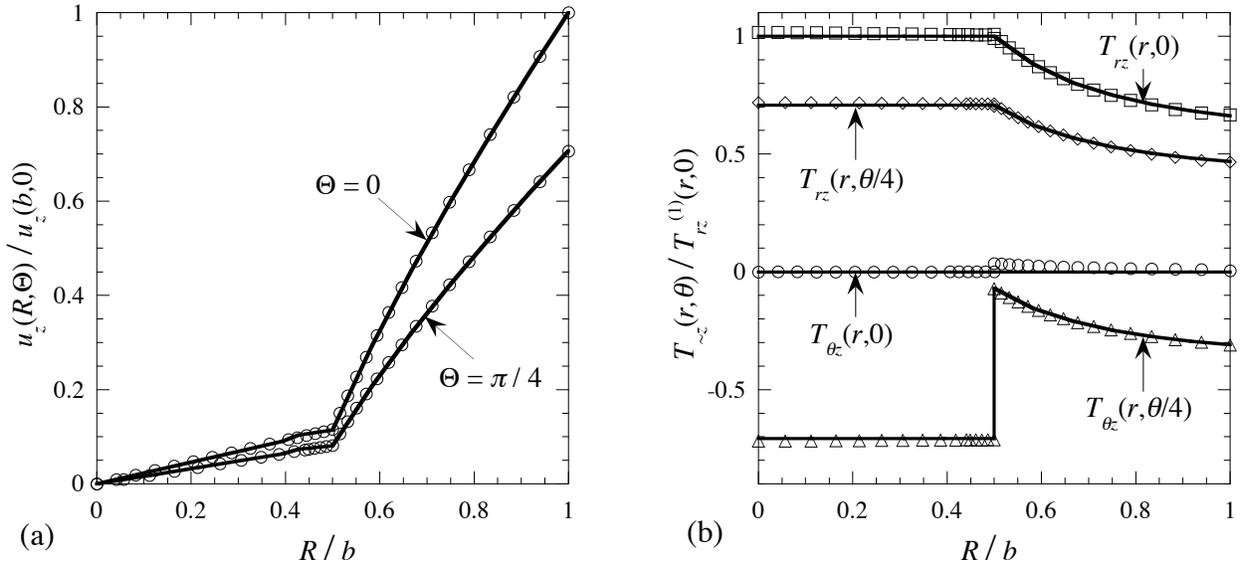}
\caption{Out-of-plane displacements (a) and shear stresses (b) as functions of $R$ along the rays $\Theta=0$ and $\pi/4$.
The continuous curves correspond to the analytic solution and the marks to the finite element simulation.}
\label{compare}
\end{center}
\end{figure}

Shown in Fig.~\ref{compare}a are the out-of-plane displacements $u_{z}=z-Z$ as functions of $R$ along the rays $\Theta=0$ and $\pi/4$.
In Fig.~\ref{compare}b the variations of the out-of-plane components of the Cauchy stress $\stress_{rz}$ and $\stress_{\theta z}$ are shown as functions of $R$ along the same rays.
In both figures the continuous curves correspond to the analytic solution and the marks to the numerical results determined at the nodes of the finite element mesh.
To minimize edge effects, the numerical results were extracted at the central cross section of the model half way between its bases.
For convenience, the displacements are normalized by the out-of-plane displacement $u_{z}(b,0)$ of the circumferential point along the ray $\Theta=0$.
The stresses are normalized by the uniform stress component $\stress_{rz}^{\rank{1}}(r,0)$ developing in the fiber along the ray $\Theta=0$.

Clearly, there is a fine agreement between the results from the analytic solution and those from the numerical simulation.
In the numerical simulations results we observe the small jump in the stress component $\stress_{\theta z}(r,0)$ at the interface between the fiber and the shell.
We verified that this jump is mesh sensitive and diminishes as the mesh becomes finer.

\section{Applications to fiber reinforced composites}

An important question concerns the ability to identify an \emph{effective} homogeneous medium whose behavior is identical to the behavior of the cylindrical composite element.
To examine this issue we follow \citet{hash&rose64jamt} and determine the traction developing at the circumferential boundary of the cylindrical composite element under the deformation (\ref{cylindrical mapping}).
From \refeq{polar nominal stress} for the nominal stress the components of the traction on the boundary are
\begineq{traction on the boundary}
t_{R}=\mu^{\rank{2}}-p^{\rank{2}},\quad
t_{\Theta}=0,\quad
t_{Z}=\gamma \tilde\mu \cos \Theta,
\eqend
where
\begineq{effective shear modulus}
\tilde\mu=\mu^{\rank{2}}\frac
{(1+\volfrac^{\rank{1}})\mu^{\rank{1}}+(1-\volfrac^{\rank{1}})\mu^{\rank{2}}}
{(1-\volfrac^{\rank{1}})\mu^{\rank{1}}+(1+\volfrac^{\rank{1}})\mu^{\rank{2}}}.
\eqend

Consider next a homogeneous cylinder made out of a neo-Hookean material with shear modulus $\mu_{0}$. 
The cylinder is subjected to the same boundary conditions as given in \refeq{boundary cylindrical mapping}.
The corresponding deformation gradient and the stress in the cylinder are constants. 
It can be easily verified that the components of the traction developing on the boundary of this cylinder are
\begineq{traction on the homogeneous}
t_{R}=\mu_{0}-p_{0},\quad
t_{\Theta}=0,\quad
t_{Z}=\gamma \mu_{0} \cos \Theta,
\eqend
where $p_{0}$ is an arbitrary pressure.

We note that if $\mu_{0}=\tilde\mu$, up to a constant arbitrary pressure, the overall out-of-plane shear response of the composite cylinder and the homogeneous cylinder are identical.
Accordingly, the out-of-plane shear behavior of a \emph{composite cylinder assemblage} \citep{hash&rose64jamt} made out of a space filling assemblage of cylindrical composite elements with various radii but with fixed ratio of the inner to the outer radii in each element will be identical to the shear response of a homogeneous neo-Hookean material with shear modulus $\tilde\mu$. 
For the idealized transversely isotropic {composite cylinder assemblage} the above homogenized result for the out-of-plane shear modulus is exact.
For more realistic microstructures with overall transverse isotropy \refeq{effective shear modulus} can be used as a straightforward and simple approximation for the out-of-plane shear modulus.

\section{Concluding remarks}
Within the framework of finite deformation elasticity, we develop a close form solution for the problem of a cylindrical composite element made out of two incompressible neo-Hookean phases subjected to out-of-plane shear loads.
We note that since the shear problem is isochoric, if the behavior of one of the two phases in the cylindrical composite element is neo-Hookean and the behavior of the second phase reduces to a form identical to that of a neo-Hookean material under shear loads (even if, in general, the second phase is compressible), the proposed solution will satisfy this problem too.
This is because under the imposed boundary conditions the deformation of the second phase must be isochoric.
We further note that the proposed solution is in agreement with the corresponding solution of \citet{hash&rose64jamt} for the analogous problem in the limit of small deformation elasticity.
We also demonstrate that the results of a corresponding finite element simulation are in agreement with the proposed solution.

In the context of fiber composites undergoing large deformations, we find that the effective out-of-plane shear behavior of a composite cylinder assemblage \citep{hash&rose64jamt} made out of two neo-Hookean phases is identical to that of a homogeneous neo-Hookean material.
An expression for the effective shear modulus of the homogeneous neo-Hookean material in terms of the properties of the two phases and their volume fractions was determined.
This expression is identical to the expression obtained by \citet{hash&rose64jamt} for the analogous problem in the limit of small deformation elasticity.
We also note that this expression is identical to the expression obtained by \citet{hash65jmps} for the bounds on the effective out-of-plane shear modulus of transversely isotropic composites in the limit of small deformation elasticity.

Finally, it is noted that this expression is also identical to an exact expression that was recently obtained by \citet{gdb05jmps} for the effective shear modulus of a special class of incompressible transversely isotropic composites undergoing finite \emph{in-plane} shear deformations.
In the limit of small deformation elasticity, the microstructure considered by \citet{gdb05jmps} is an optimal microstructure attaining the Hashin-Shtrikman bounds of \citet{hash65jmps} on the in-plane shear modulus for the class of incompressible transversely isotropic composites.
Thus, we observe an interesting analogy where in 
two works dealing with transversely isotropic composites undergoing finite in-plane and out-of-plane shear deformations, the expressions for the effective shear moduli are identical to each other and to the ones previously obtained in the limit of small deformation elasticity.

\begin{small}

\end{small}
%
%
\end{document}